# Proton Extraction from IHEP Accelerator Using Bent Crystals

A.G. Afonin, V.T. Baranov, V.M. Biryukov[*], V.N. Chepegin, Yu.A. Chesnokov, Yu.S. Fedotov,
A.A. Kardash, V.I. Kotov, V.A. Maisheev, V.I.Terekhov, E.F. Troyanov

*Institute for High Energy Physics, 142281 Protvino, Russia*

**Abstract.** IHEP Protvino has pioneered the wide practical use of bent crystals as optical elements in high-energy beams for beam extraction and deflection on permanent basis since 1989. In the course of IHEP experiments, crystal channeling has been developed into efficient instrument for particle steering at accelerators, working in predictable, reliable manner with beams of very high intensity over years. Crystal systems extract 70 GeV protons from IHEP main ring with efficiency of 85% at intensity of $10^{12}$, basing on multi-pass mechanism of channeling proposed theoretically and realised experimentally at IHEP. Today, six locations on the IHEP 70-GeV main ring of the accelerator facility are equipped by crystal extraction systems, serving mostly for routine applications rather than for research and allowing a simultaneous run of several particle physics experiments, thus significantly enriching the IHEP physics program. The long successful history of large-scale crystal exploitation at IHEP should help to incorporate channeling crystals into accelerator systems worldwide in order to create unique systems for beam delivery. We report recent results from the research and exploitation of crystal extraction systems at IHEP.

## 1. Introduction

The idea to deflect proton beams using bent crystals, originally proposed by Edouard Tsyganov in 1976 [1], was demonstrated shortly later, in 1979, by a Soviet-American team in Dubna on proton beams of a few GeV energy [2]. In the following, the physics of bent crystal channeling has been studied at many high energy labs; see e.g. refs. [3,4] for reviews.

Crystal-assisted extraction from accelerator was demonstrated for the first time in 1984 in Dubna [5] at beam energies of 4-8 GeV and then tested at IHEP in Protvino starting from 1989 by exposing, firstly, a silicon crystal bent by 85 mrad to the 70 GeV proton beam of U-70 [6,7]. The Protvino experiment eventually pioneered the first regular application of crystals for beam extraction: the Si crystal, originally installed in the vacuum chamber of U-70, served without replacement over 10 years, delivering beam for particle physicists all this time. However, its channeling efficiency was a small fraction of percent.

In the 1990's, an important milestone was obtained at the CERN SPS. Protons diffusing from a 120 GeV beam were extracted at an angle of 8.5 mrad with a bent silicon crystal. Efficiencies of ~10-20%, orders of magnitude higher than the values achieved previously, were measured for the first time [8-10]. The extraction studies at SPS clarified several aspects of the technique. In addition, the extraction results were found in fair agreement with Monte Carlo predictions [8,11-13]. In the late 1990's, another success came from the Tevatron extraction experiment where a crystal was channeling a 900-GeV proton beam with an efficiency of ~30% [14-16]. The simulation [17] predicted the efficiency of 35% for a realistic crystal in the Tevatron experiment. During the FNAL test, the halo created by beam-beam interaction in the periphery of the circulating beam was extracted from the beam pipe without measurable effect on the background seen by the experimental detectors.

---

[*] Corresponding author. E-mail: biryukov@mx.ihep.su  Web: http://www1.ihep.su/~biryukov/

It was predicted [18,19] that efficiency of crystal channeling extraction can be boosted to much higher values by multiple particle encounters with a shorter crystal installed in a circulating beam. The existence of multi-pass mechanism was confirmed in early experiments at CERN SPS and Tevatron [9,15]. To clarify this mechanism a new experiment was started at IHEP at the end of 1997, with intention to test very short crystals and achieve very high efficiencies of extraction [20-22].

## 2. High efficiency channeling

In order to let the circulating particles encounter the crystal many times and suffer less scattering and nuclear interactions in the crystal, one has to minimise the crystal length down to some limit set by the physics of channeling in a strongly bent crystal [12]. This optimisation was studied in Monte Carlo simulations in general and for the experiments at CERN SPS and the Tevatron [11-13,17], taking into account the circulation of particle in the accelerator ring over many turns with multiple encounters with a bent crystal. On every encounter with a crystal, particle is tracked through the curved crystal lattice by the simulation (*ab initio*) code described in refs. [23-26]. The predictions of this code were earlier successfully tested for variety of channeling-related phenomena like feed-out [27] and feed-in [28], energy losses [29], and in particular for crystal extraction at CERN SPS [8,11,12], Tevatron [15], RHIC [30-32], and IHEP U-70 [20-22,33-37].

Monte Carlo study done for the experiment at IHEP has predicted [37] that a crystal can be shorten very much, down to ~1 mm along the 70-GeV beam in the extraction set-up of IHEP U-70. The benefits from this optimisation were expected enormous: the crystal extraction efficiency could be as high as over 90%.

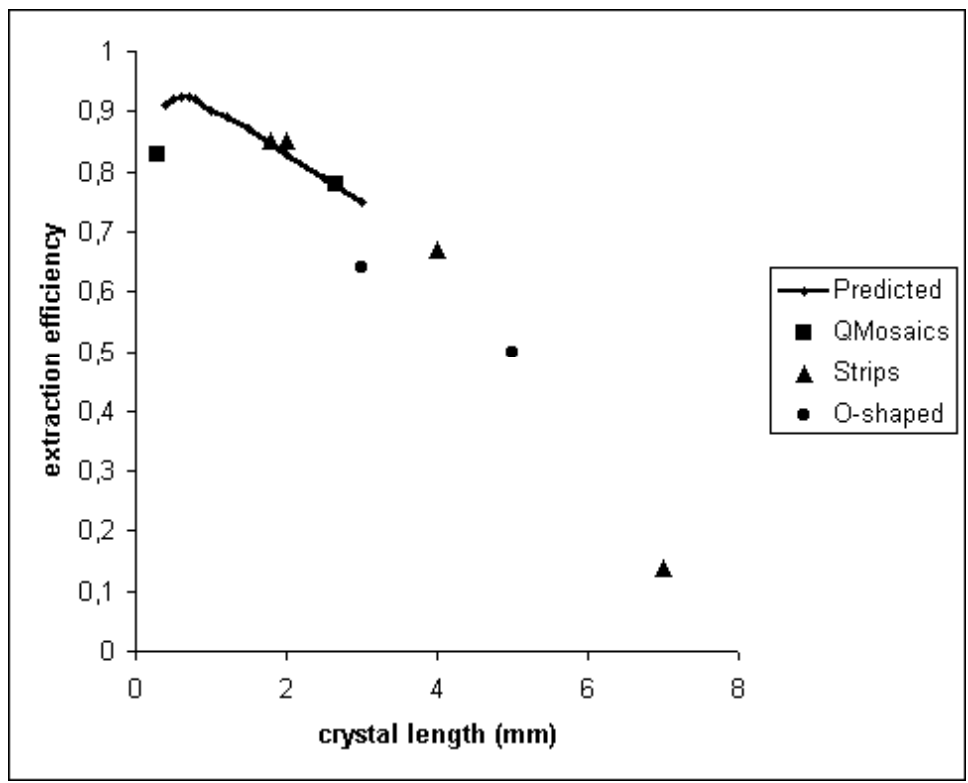

**Fig. 1** Crystal extraction efficiency measured for 70 GeV protons as a function of the crystal length. IHEP measurements [-] for three types of crystal deflectors - O-shaped (circles), strips (triangles) and quasi-mosaics (squares), - and earlier Monte Carlo prediction [21].

The effect of shortening the crystal on the efficiency of multi-turn extraction is explained by Figure 1, where shown are both the predicted [37] dependence of the IHEP crystal extraction efficiency as a function of the crystal length, and some history of the measurements since 1997 [20-22,33-37]. Here, two very fresh and very interesting experimental points (squares) [38] have been added to the plot, obtained with a new bending technique of quasi-mosaic crystals. The shortest crystal used at IHEP is now just 0.3 mm along the beam (!) still showing the efficiency of more than 80%. Notice that the experimental data plotted is obtained with different bending angles in the range 0.5-2.3 mrad and the predicted curve is for 0.9 mrad, so the plot is to show the global trend first of all. Some more detail on crystal size and bending is given in Table 1.

**Table 1**

| № | location | type | bend angle, mrad | size l×h×R, mm | efficiency, % | energy, GeV | extraction scheme |
|---|---|---|---|---|---|---|---|
| 1 | SS-106 | S | 1.0 | 2.0×35×0.5 | 85<br>80 | 70<br>70 | 106-24-26<br>106-20-22-26 |
| 2 | SS-106 | O | 0.7 | 3.5×5.0×0.7 | 60 | 70 | 106-24-26 |
| 3 | SS-19 | S | 2.0 | 5.0×45×0.5 | 67 | 70 | 20-22-26 |
| 4 | SS-19 | O | 2.1 | 5.0×5.0×0.7 | 65 | 70 | 20-22-26 |
| 5 | Bl. 22 | S | 0.8 | 1.9×45×0.5 | 85 | 70 | 24-26 |
| 6 | Bl. 22 | S | 0.9 | 1.8×45×0.5 | 80 | 50 | 24-26 |

Producing bent crystal deflectors of required size and curvature is not an easy task, moreover as one takes into account that deflector has to be placed in a circulating beam and any extra disturbance to halo particles must be avoided. Excellent crystalline deflectors used at IHEP were produced at IHEP and PNPI in different approaches. In PNPI, the technique of "O"-shaped crystals of small size (Fig. 2) was applied [20]. In IHEP, "strip" type of crystal deflector was invented (Fig. 2), beneficially using the effect of anticlastic bending [33]. Some of the silicon wafers used for deflectors have undergone chemical polishing treatment in Ferrara [39]. Three different "strip"-type crystals (bent in IHEP from different silicon wafers) have demonstrated record 85% efficiency of channeling [33,39]. New approaches to crystal bending are being studied.

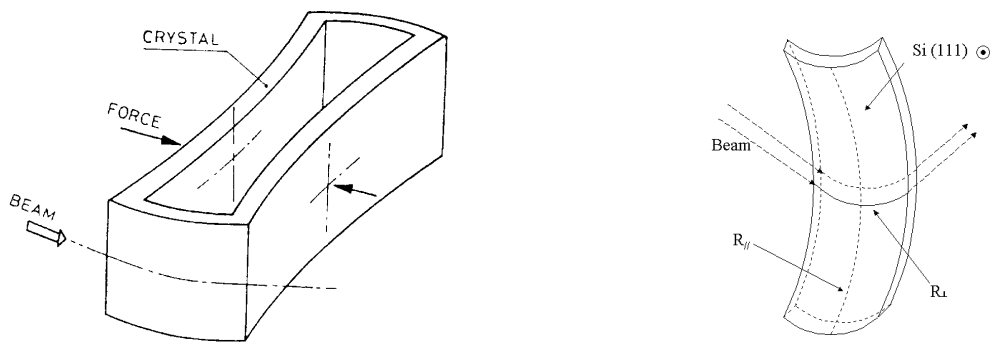

**Fig. 2** Two examples of IHEP crystals with bending schemes. Left: O-shaped crystal as used in IHEP and RHIC. Right: Strip-type crystal used in IHEP.

The experimentally recorded high efficiency followed nicely the prediction as seen in Fig. 1. Compared to the CERN SPS and Tevatron experiments, the efficiency is improved by a factor of 3-6 while the crystal size along the beam was reduced by a factor of 15-20 (from 30-40 mm to less than 2 mm).

Experimentally, the extraction efficiency was defined as the ratio of the extracted beam intensity as measured in the external beam line to all the beam loss in the circulating beam. Each measurement included the statistics from several hundred machine cycles. A remarkable feature of the IHEP extraction is that the record high efficiency of about 85% is obtained even when the entire beam stored in the ring is dumped onto the crystal. Typically, crystal channeled ~$10^{12}$ protons (up to $3 \cdot 10^{12}$ in some runs) in a spill of 0.5-1 s duration.

**Figure 3 a**

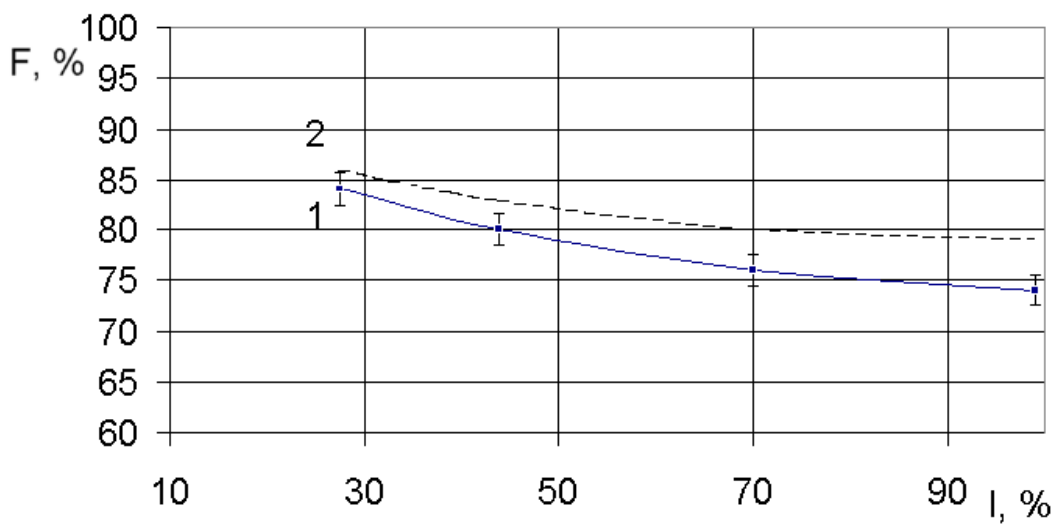

**Figure 3 b**

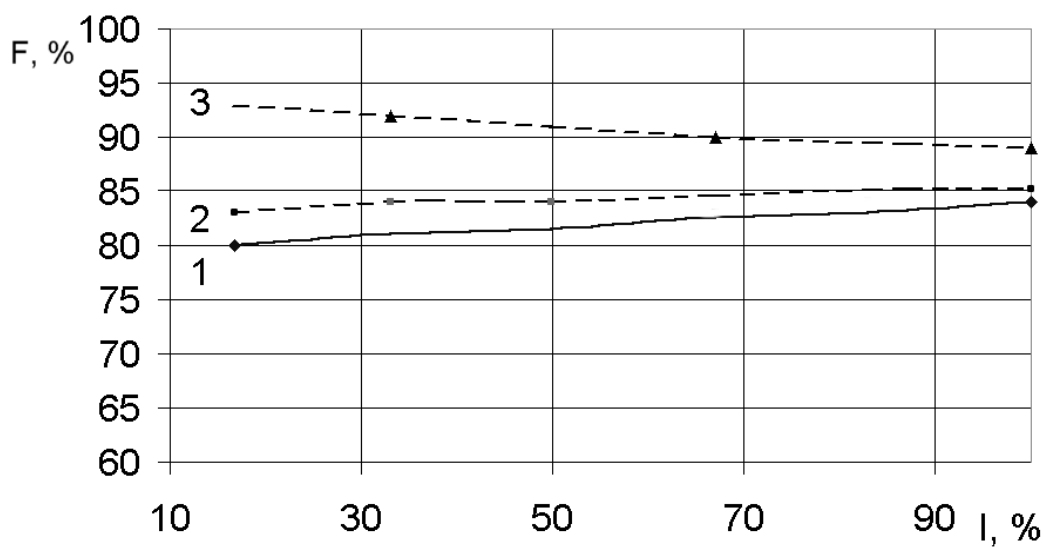

**Fig. 3** Crystal extraction efficiency at 70 GeV as a function of the beam store fraction dumped onto the crystal. Measured (solid lines, 1) on two locations (a, b) in the U-70 ring and simulated (dashed, 2). Curve 3 is simulation for the same crystal installed closer to the orbit.

Figure 3 shows how the efficiency changed versus the fraction of the beam store used in the channeling experiment. Two different locations on the U-70 ring were used in the data presented in Fig. 3, straight section 106 (Figure 3a) and block 22 (Figure 3b), and similar crystals (strip-type) employed. Different slopes of the experimental dependences are due to the drift of the mean incident angle when a beam is being dumped onto the crystal. The solid lines (1) in Fig. 3 a and b represent the measured data while the dashed lines (2) are the Monte Carlo results for the absolute figure of efficiency of extraction of 70 GeV protons on these two locations. The results of theory and experiment are in fair agreement.

Notice that the experimental figure of efficiency attributes all the observed beam losses to the crystal alone while actually there could be additional losses of the channeled beam when it is transferred along the accelerator extraction line, for instance on the septum magnet partition wall, before the extracted beam is delivered to the external line where its intensity is measured and used for evaluation of the extraction efficiency.

Further on in simulations for the location of block 22 we tried to optimize crystal radial position with respect to the beam orbit and to the edge of the septum magnet downstream. With crystal installed closer to the orbit, the efficiency might be improved further, as shows line (3) in Fig. 3b, and be about 90% with the same crystal.

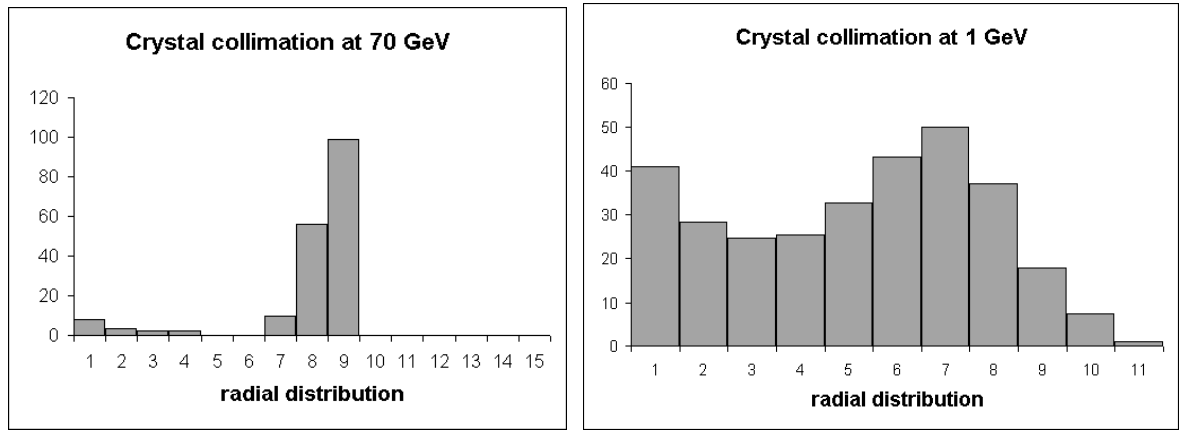

**Fig. 4** The radial beam profile observed 20 m downstream of the channeling crystal, at top energy (70 GeV, left) and at injection plateau (1.3 GeV, right); the crystal is the same. The major peak on each profile is channeled particles.

IHEP has many locations on the U-70 ring where crystals are installed for extraction and collimation studies. On two of these locations (dedicated for crystal collimation) the channeled beam is not extracted from the ring but is intercepted by a collimator downstream. In a collimation experiment, a bent crystal is positioned upstream of a secondary collimator (stainless steel absorber 4 cm wide, 18 cm high, 250 cm long) and closer to the beam in the horizontal plane. The profilemeter records the radial distribution of the particles incident on the entry face of the secondary collimator (Fig. 4). This distribution includes the peak of channeled particles deflected into the depth of the collimator, and the nonchanneled multiply scattered particles peaked at the edge of the collimator. The efficiency figures as measured on the extraction set-up were reproduced on the collimation set-up where the intensity of the channeled beam is obtained by integration of the peak in the profile. This set-up allows an independent check of the crystal channeling efficiencies, and also gives opportunity to work with different bending angles unlike in the extraction set-up where crystal bending angles are more dictated by the geometry of extraction.

The channeling experiment was repeated at the injection plateau of the U-70, with 1.3 GeV protons, on the same collimation set-up with the same crystal. As one can see in Figure 4 the channeling effect is still quite profound although the energy was lowered by two orders of magnitude.

## 3. Crystal channeling at a ramping energy

In the recent IHEP experiment channeling was tested on the same crystal collimation set-up in a broad energy range made available in the main ring of U-70 accelerator. Earlier, the experiment was performed at the top energy flattop, 70 GeV, and at the injection flattop, 1.3 GeV, of U-70 machine. This time the tests were done at seven intermediate energies and, importantly, it was not possible to arrange a flattop for each energy. During the acceleration part of the machine cycle, on a certain moment corresponding to the energy of the test, the beam was dumped in a short time onto the crystal.

These measurements are summarized in Figure 5 showing the ratio of the channeled particles to the entire beam dump (crystal channeling efficiency) as measured and as predicted by Monte Carlo simulation. Figure 6 shows the examples of the radial beam profile observed at the entry face of the collimator with crystal working at 12 GeV and at 45 GeV. One can see that the same crystal shows efficient work from injection through the ramping up to the top energy.

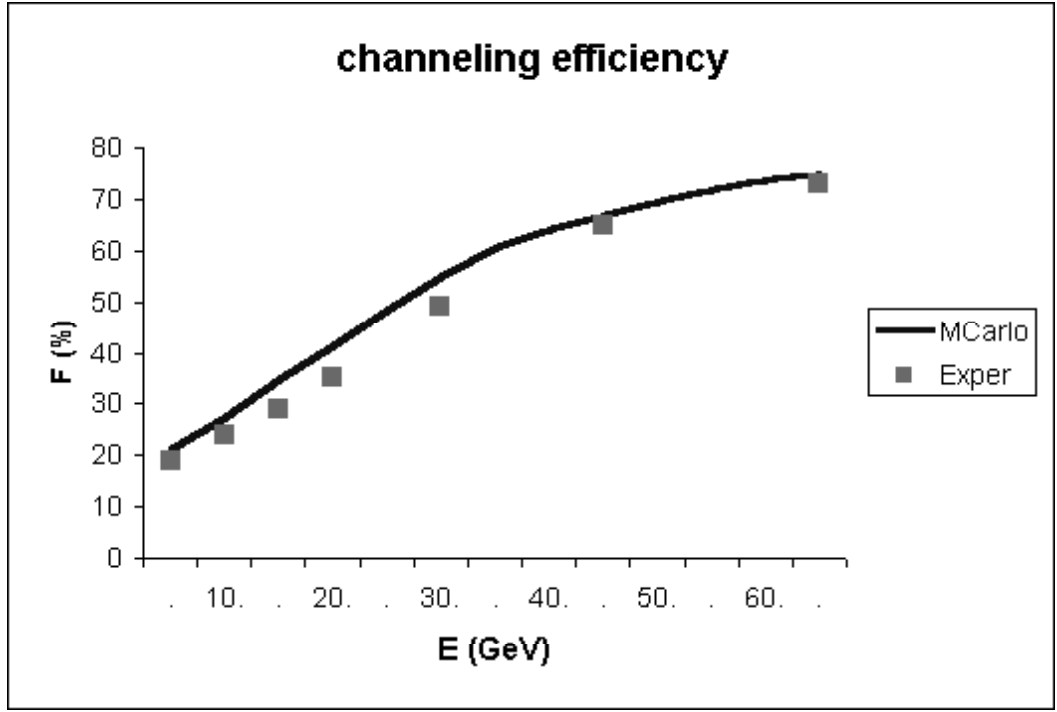

**Fig. 5** Crystal efficiency as measured and as expected from simulation. The case of ramping energy in U-70.

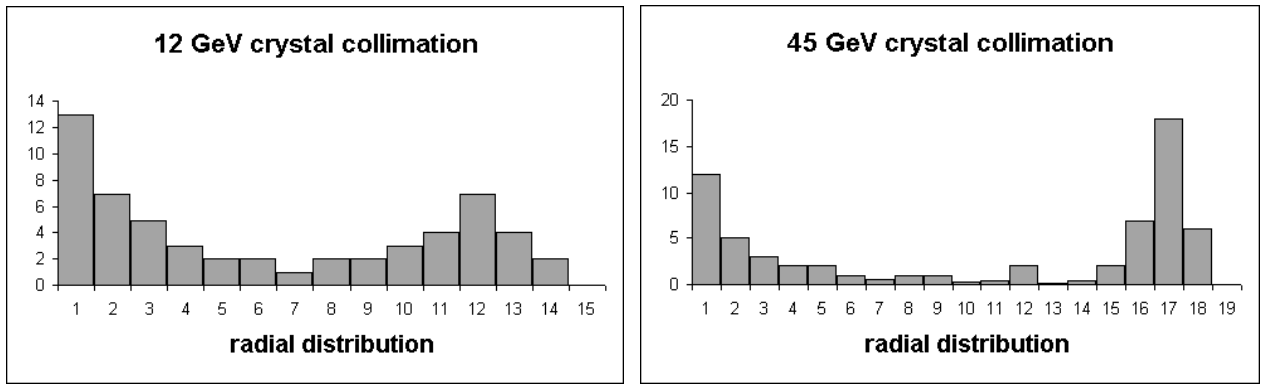

**Fig. 6** The same as in Fig. 4 but at 12 GeV (left) and at 45 GeV (right).

## 3. Beam delivery to particle physics experiments

Crystal bending of a particle beam at a small angle (1-2 mrad) is not sufficient for direct extraction of particles out of the accelerator ring because the straight sections are short. Therefore, one had to use several septum-magnets of the existing extraction system. As the main experimental set-ups are connected to this extraction system, the beam extraction by means of crystal could be used for carrying out a number of physical experiments. On different locations of the accelerator, three similar crystal stations $Si_{19}$, $Si_{22}$ and $Si_{106}$, were installed as shown in Fig. 7. Here, the index notes the number of straight section or the number of magnet block where crystals are installed. Each station can ensure the extraction of proton beam towards the existing extraction line provided that corresponding combination of local orbit distortion is involved.

In addition to these stations, three more stations: $Si_{30}$, $Si_{84}$ and $Si_{86}$ have been installed (Fig. 7). Crystal station $Si_{30}$ serves for splitting a small part (~$10^7$) of the beam extracted towards the beamline № 8 and deflecting it at ~9 mrad towards the beamline № 22. Crystal stations $Si_{84}$ and $Si_{86}$ are situated in a test zone and are dedicated for crystal tests before their installation into work stations $Si_{19}$, $Si_{22}$ and $Si_{106}$, and also for channeling research, in particular for the studies of beam collimation regimes using bent crystals.

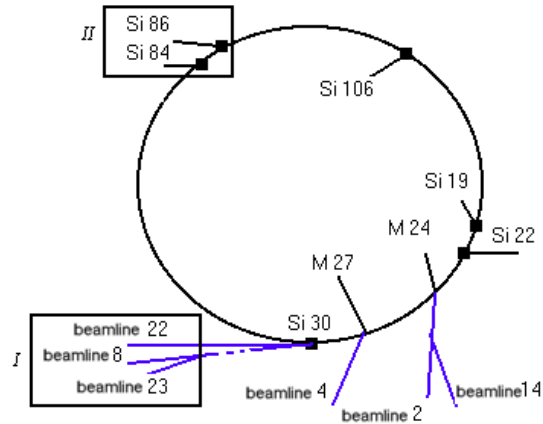

**Fig. 7** Beam extraction schemes at U-70. Crystal stations Si 19, Si 22, Si 30, Si 84, Si 106 and internal targets M24 and M27. The zone of experimental set-ups (*I*) and crystal test zone (*II*). Other numbers (2, 4, 8, 14, 22, 23) indicate beamlines.

Let us look at the chosen schemes of extraction closer. Fig. 8 shows several schemes used for beam extraction. In the first scheme, crystal station $Si_{19}$ with two crystals (one of them spare) is installed in the straight section № 19. It provides an independent translation of each crystal in radial plane and the change of crystal orientation. The circulating beam is brought to the crystal by the method of local beam distortion (magnet blocks № 15/21 and № 16/22). The particles impinging on crystal and trapped into channeling mode are bent by crystal at ~ 1.7÷2.5 mrad (depending on the bending angle of the crystal used) and enter the aperture of septum magnet OM20, avoiding the septum partition wall. After bending in septum magnets OM22 and OM26 (curve 2 in Fig. 8), the particles exit the accelerator vacuum chamber in straight section 30. This scheme is attractive also because it allowed to organize a simultaneous extraction of beam by crystal and by two internal targets $M_{24}$ and $M_{27}$ (curve 1 in Fig. 8).

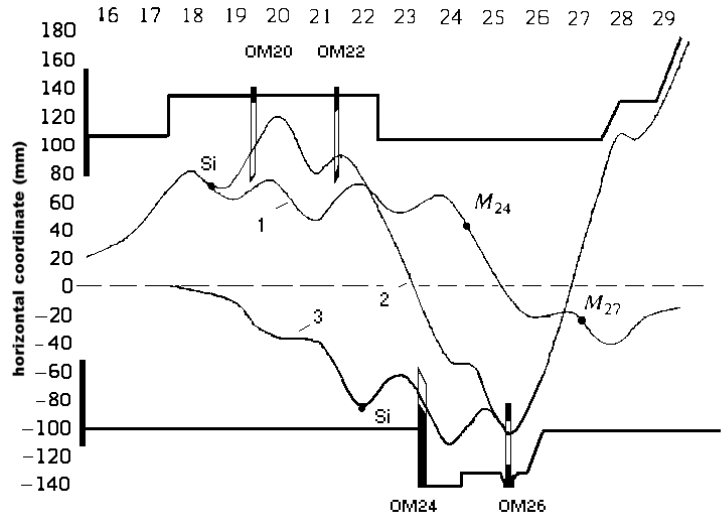

**Fig. 8** Examples of the used schemes of beam extraction, with crystal stations indicated (Si). Details in text.

When using another crystal station $Si_{22}$, positioned in the middle of the magnet block № 22, proton extraction was realized through the septum magnets OM24 and OM26 (curve 3 on Fig. 8). In this case, two pairs of magnet blocks: 20/26 and 18/30 are used to bring beam onto the crystal. Working with third station $Si_{106}$, installed in the straight section № 106 (not shown in Fig. 8), also two pairs of magnet blocks: 103/109 and 104/110 are used for beam steering onto crystal, and the extraction was realized in two ways: through septum magnets OM20, 22 and 26, and through septum magnets OM24 and OM26. Therefore, the third scheme can replace any of the first two ones in case of troubles.

Figure 9 illustrates how stable is the work of our crystal extraction system; this data is for extraction at 50 GeV where crystal efficiency was 80% (a bit lower than at 70 GeV in the same set-up). Figure 9a shows the intensity of the extracted beam per cycle during 180 cycles as measured in the external beamline. Figure 9b shows the efficiency of crystal extraction measured for each cycle of those 180 ones. The system works stable enough at the intensity of the extracted beam of $10^{12}$ proton per cycle. Our "strip" crystals worked in this regime over 3000 hours in recent accelerator runs.

Crystal-assisted extraction of protons allows in principle a simultaneous work of several internal targets as confirmed experimentally in 1991 at the IHEP U-70 accelerator [40]. Realization of this regime with use of short crystals opens the possibility of simultaneous work of several experimental set-ups during all the flattop of the accelerator magnet cycle, which leads to significant economy. Notice that a classical resonance slow extraction is not compatible with a parallel work of internal targets. In this regime, the "Complex of Tagged Neutrinos" set-up has been working. This physics experiment needed a high intensity of extracted beam with a longest possible duration. These requirements could not be ensured earlier, before the use of crystal extraction regime. The experiment studied three-particle decays: $K^+ \to \pi^+ \pi^0 \pi^0$, $K^- \to \pi^- \pi^0 \pi^0$. As the yield of $K^+$-mesons is significantly higher than that of $K^-$-mesons, with a change of regime one had to increase/reduce proportionally the intensity of extracted proton beam.

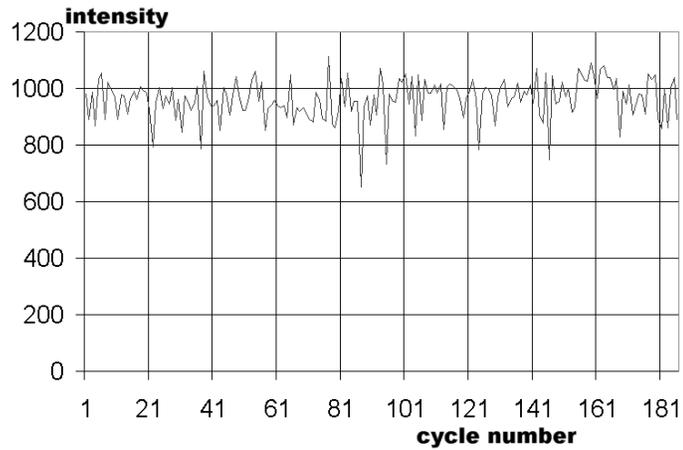

**Fig. 9** (a) Intensity of extracted beam per cycle measured over 180 cycles in external beamline.

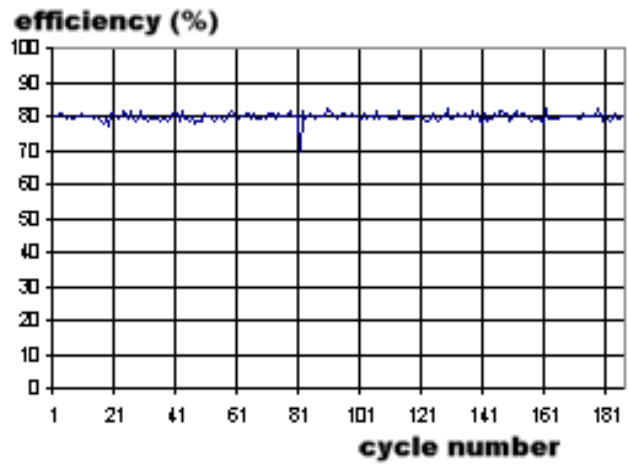

**Fig. 9** (b) Efficiency of crystal extraction measured for each cycle of 180 cycles.

One of most important parameters of extracted beams, from viewpoint of physics experiments, is the time structure of beam or effective time of extraction. In the course of experiment, minimal change of the particle rate is preferred. In the considered regime, the time structure was suitable for all simultaneously working experimental set-ups. The effective time of extraction here reached 98%. The characteristics of crystal delivered beams satisfied the physicists over such a range of parameters as intensity, duration, beam quality, beam size on a target.

One peculiarity of crystal work in parallel to internal targets is the significant increase in the size of the circulating beam due to scattering on internal targets. This effect leads to existence of two mechanisms that bring particles to crystals. One is local beam distortion. Another one is the scattering on internal targets. This latter mechanism we used for crystal extraction of low-intensity beam. The uniform time structure of extraction was defined by uniform dump of the circulating beam on internal targets. Here, the intensity of the channeled beam reached $10^6$ proton/cycle.

The crystal-assisted extraction scheme developed and invented at our accelerator opens way for further modifications. In particular, in order to increase the number of experiments simultaneously carried out at U-70 accelerator we installed in the straight section 30 another station of crystal deflectors with a crystal bent 9 mrad in the direction of beamline 22. This crystal deflected a small fraction (~$10^7$ proton/cycle) of the beam already extracted towards the beamline 23 and delivered this fraction to another set-up situated on beamline 23.

As the result of introducing this regime, we managed to increase the number of experimental set-ups working simultaneously during the flattop of magnet cycle up to five, without taking into account the possibility of a parallel "shadow" work of yet another two internal targets. This scheme of work is most efficient when the main experimental set-ups do not plan essential changes of the regime. In the perspective of work with 50 GeV proton beam, a possibility appears to increase the flattop duration to 4 s. In this case, the application of the developed crystal extraction schemes seems most adequate and efficient. The first such experiment was carried out in 2003. In this experiment, done at 50 GeV, crystal extraction with duration of 3.2 s was obtained.

The considered regimes of extraction system with assistance of bent crystals essentially increase the number of simultaneously working experimental set-ups and improve their data-taking rate thanks to the increase in duration and intensity of the extracted beam. Since 1999, the use of bent crystals for ensuring the high energy physics program became regular in all accelerator runs of U-70. About one half of all extracted beams with intensities up to $10^{12}$ at IHEP are now obtained with channeling crystals.

## 4. Conclusion

The IHEP Protvino 70 GeV accelerator facility is largely "crystallized" and became a "fabric" of channeled beams. Thanks to crystals, up to five physics experiments can be run simultaneously at IHEP. Crystal extraction efficiency of 85% is routinely obtained for a beam of $10^{12}$ protons at 70 GeV; three different crystals of the same size and geometry have shown the same efficiency, in excellent agreement with Monte Carlo prediction from first principles. The same crystal works efficiently over full energy range, from injection flattop through ramping up to top energy plateau, as demonstrated experimentally from 1 through 70 GeV. At the above said intensity, there were no problems with high intensity or lifetime of a crystal. One of the crystals has served for extraction over 10 years, from 1989 to 1999, without replacement. Many types of deflectors were tested and found successful. We presently work on bringing the regular intensity of channeled beams up to $10^{13}$ per spill.

The reported experience of IHEP makes a strong case for application of crystal-based technique worldwide. One opportunity is crystal steering of a Large Hadron Collider beam [41-43] with the purpose of collimation or extraction, where crystal should be very efficient according to simulations [25,41]. Our Monte Carlo model successfully predicts crystal work in the circulating beam, as observed in crystal extraction experiments at IHEP, CERN SPS, RHIC, and Tevatron at energies up to 900 GeV. Crystal channeling deflection, extraction and collimation are now established beam instruments.

**Acknowledgement**. This work was supported by INTAS-CERN grants 132-2000 and 03-52-6155.

## References

[1] E.N. Tsyganov. Fermilab preprints TM-682, TM-684 (1976)
[2] A.S. Vodopianov et al. JETP Lett. **30** (1979) 474
[3] R.A. Carrigan and J. Ellison, eds. Relativistic Channeling (NY: Plenum, 1987)
[4] V.M.Biryukov, Yu.A.Chesnokov, V.I.Kotov. Crystal Channeling and its Application at High Energy Accelerators. (Springer, Berlin: 1997)
[5] V.V. Avdeichikov et al. JINR Rapid Comm. 1-84 (1984) 3.
[6] A. A. Asseev, M. D. Bavizhev, E. A. Ludmirsky, V. A. Maisheev and Yu. S. Fedotov, Nucl. Instrum. Methods Phys. Res., Sect. A **309** (1991) 1.
[7] A. A. Asseev, E. A. Myae, S. V. Sokolov and Yu. S. Fedotov, . Nucl. Instrum. Methods Phys. Res., Sect. A **324** (1993) 31
[8] H. Akbari et al. Phys. Lett. B **313** (1993) 491
[9] X. Altuna et al., Phys. Lett. B **357** (1995) 671
[10] A. Baurichter, C. Biino, M. Clément, N. Doble, K. Elsener, G. Fidecaro, A. Freund, L. Gatignon, P. Grafström, M. Gyr *et al.* Nucl.Instrum.Meth.B**164-165**: (2000) 27-43
[11] V. Biryukov. EPAC 1994 Proceedings (London), p.2391. "Simulation of the CERN-SPS Crystal Extraction Experiment"
[12] V. Biryukov. Nucl. Instr. and Meth. B **117** (1996) 463

[13] V. Biryukov. EPAC 1998 Proceedings (Stockholm), p.2091. "Analytical Theory of Crystal Extraction".
[14] C. T. Murphy, R. Carrigan, D. Chen, G. Jackson, N. Mokhov, H. -J. Shih, B. Cox, V. Golovatyuk, A. McManus, A. Bogacz *et al*., *Nucl. Instrum. Meth.* B **119** (1996) 231
[15] R. A. Carrigan et al., Phys. Rev. ST Accel. Beams **1** (1998) 022801.
[16] R.A. Carrigan et al. Phys. Rev. ST Accel. Beams **5** (2002) 043501.
[17] V. Biryukov, Phys. Rev. E **52** (1995) 6818
[18] V. Biryukov, Nucl. Instrum. Methods Phys. Res., Sect. B **53**, 202 (1991).
[19] A. M. Taratin, S. A. Vorobiev, M. D. Bavizhev and I. A. Yazynin, Nucl. Instrum. Methods Phys. Res., Sect. B **58**, 103 (1991).
[20] A.G. Afonin et al. Phys. Lett. B **435** (1998) 240-244.
[21] A.G. Afonin et al. JETP Lett. **67** (1998) 781-785.
[22] A.G. Afonin et al. JETP Lett. **68** (1998) 568-572.
[23] V. Biryukov. CERN SL Note 93-74 AP (1993). ). "CATCH 1.4 User's Guide".
[24] V. Biryukov. Phys. Rev. E **51** (1995) 3522.
[25] V. Biryukov. Phys. Rev. Lett. **74** (1995) 2471.
[26] V. Biryukov. Phys. Rev. E **52** (1995) 2045.
[27] V. M. Biryukov, Yu. A. Chesnokov, N. A. Galyaev, V. I. Kotov, I. V. Narsky, S. V. Tsarik, V. N. Zapolsky, O. L. Fedin, M. A. Gordeeva, Yu. P. Platonov and A. I. Smirnov, Nucl. Instrum. Meth. B **86** (1994) 245
[28] V.M. Biryukov, V.I. Kotov, Y.A. Chesnokov. Physics-Uspekhi **37** (1994) 937
[29] S. Pape Moller, V. Biryukov, S. Datz, P. Grafstrom, H. Knudsen, H.F. Krause, C. Scheidenberger, U.I. Uggerhoj, C.R. Vane. Phys. Rev. A **64** (2001) 032902
[30] D. Trbojevic, V. Biryukov, M. Harrison, B. Parker, P. Thompson, A. Stevens, N. Mokhov, A. Drozhdin. EPAC 1998 Proceedings (Stockholm), p.2146. "A Study of RHIC Crystal Collimation"
[31] R.P. Fliller III, A. Drees, D. Gassner, L. Hammons, G. McIntyre, S. Peggs, D. Trbojevic, V. Biryukov, Y. Chesnokov, V. Terekhov AIP Conf. Proc. **693** (2003) 192-195
[32] R.P. Fliller III, A. Drees, D. Gassner, L. Hammons, G. McIntyre, S. Peggs, D. Trbojevic, V. Biryukov, Y. Chesnokov, V. Terekhov. Nucl. Instrum. Methods Phys. Res., Sect. B, in press
[33] A.G. Afonin et al., Phys. Rev. Lett. **87**, 094802 (2001).
[34] A.A. Arkhipenko et al. Instrum. Exp. Tech. **43** (2000) 11-15.
[35] A.G. Afonin et al. JETP Lett. **74** (2001) 55-58
[36] A.G. Afonin et al. Instrum. Exp. Tech. **45**(4) (2002) 476
[37] V.I. Kotov et al. EPAC 2000 Proceedings (Vienna), p.364. CERN-LHC-2000-007-MMS
[38] V.T. Baranov et al., to be published.
[39] V.M. Biryukov, et al. Rev. Sci. Instrum. **73** (2002) 3170-3173
[40] A.A. Asseev et al. IHEP Preprint 91-46 (Protvino, 1991)
[41] V.M. Biryukov, V.N. Chepegin, Yu.A. Chesnokov, V. Guidi, W. Scandale. Nucl. Instrum. Methods Phys. Res., Sect. B, in press. V.M. Biryukov [ArXiv:physics/0307027].
[42] U.I. Uggerhoj and E. Uggerhoj. Nucl. Instrum. Methods Phys. Res., Sect. B, in press
[43] V.M. Biryukov, "In-Situ Calibration Of Hadron Forward Calorimeters", talk given at the 9[th] RDMS CMS conference (Minsk, 2004), http://agenda.cern.ch/fullAgenda.php?ida=a044186